\documentclass[12pt,a4paper]{article}
\usepackage{amsmath}
\usepackage{amssymb}
\usepackage{hyperref}
\usepackage[height=8.8in,width=6.45in]{geometry}

\numberwithin{equation}{section}

\usepackage{tikz}
\usetikzlibrary{arrows}
\tikzset{>=latex}

\def\fnote#1#2{\begingroup\def\thefootnote{#1}\footnote{#2}\addtocounter
{footnote}{-1}\endgroup}

\def\cN{\mathcal{N}{}}

\def\SU{\mathop{\mathrm{SU}}\nolimits}

\newcommand{\sll}{\mathrm{sl}}
\newcommand{\ZZ}{\mathbb{Z}}
\newcommand{\SL}{\mathrm{SL}}
\newcommand{\RR}{\mathbb{R}}
\newcommand{\U}{\mathrm{U}}
\newcommand{\ul}{\mathrm{u}}

\newdimen\tableauside\tableauside=1.0ex
\newdimen\tableaurule\tableaurule=0.4pt
\newdimen\tableaustep
\def\phantomhrule#1{\hbox{\vbox to0pt{\hrule height\tableaurule
width#1\vss}}}
\def\phantomvrule#1{\vbox{\hbox to0pt{\vrule width\tableaurule
height#1\hss}}}
\def\sqr{\vbox{%
  \phantomhrule\tableaustep

\hbox{\phantomvrule\tableaustep\kern\tableaustep\phantomvrule\tableaustep}%
  \hbox{\vbox{\phantomhrule\tableauside}\kern-\tableaurule}}}
\def\squares#1{\hbox{\count0=#1\noindent\loop\sqr
  \advance\count0 by-1 \ifnum\count0>0\repeat}}
\def\tableau#1{\vcenter{\offinterlineskip
  \tableaustep=\tableauside\advance\tableaustep by-\tableaurule
  \kern\normallineskip\hbox
    {\kern\normallineskip\vbox
      {\gettableau#1 0 }%
     \kern\normallineskip\kern\tableaurule}%
  \kern\normallineskip\kern\tableaurule}}
\def\gettableau#1 {\ifnum#1=0\let\next=\null\else
  \squares{#1}\let\next=\gettableau\fi\next}

\begin{document}

\begin{titlepage}

\begin{flushright}
FIAN/TD-23/12
\end{flushright}
\vskip 2cm

\begin{center}
{\Large \bfseries
Conformal blocks of Chiral fields in $\cN=2$ SUSY CFT 
\\[0.7em]
and	Affine Laumon Spaces
}

\vskip 1.2cm

{V. Belavin$^{1,2}${\,}\fnote{*}{E-mail: belavin@lpi.ru}}

\bigskip
\bigskip

\begin{tabular}{ll}
$^{1}$~\parbox[t]{0.9\textwidth}{\normalsize\raggedright
{\it Theoretical Department, Lebedev Physical Institute, RAS, Moscow, Russia}}\\
$^{2}$~\parbox[t]{0.9\textwidth}{\normalsize\raggedright
{\it Institute for Information Transmission Problems, RAS, Moscow, Russia}}
\end{tabular}

\vskip 2.5cm

\textbf{Abstract}
\end{center}

\medskip
\noindent
We consider the problem of computing $\cN=2$ superconformal block functions.
We argue that the Kazama-Suzuki coset realization of $\cN=2$ superconformal algebra in terms of 
the affine $\widehat{\sll}(2)$ algebra provides relations between
 $\cN=2$ and $\widehat{\sll}(2)$ conformal blocks. We show that for $\cN=2$ chiral fields 
the corresponding $\widehat{\sll}(2)$ construction of the conformal blocks is based on the 
ordinary highest weight representation. We use an AGT-type correspondence to relate the four-point
$\widehat{\sll}(2)$ conformal block with Nekrasov's 
instanton partition functions of a four-dimensional $\cN=2$ $SU(2)$ gauge theory in the presence 
of a surface operator. Since the previous relation proposed by Alday and Tachikawa 
requires some special modification of the conformal block function, we revisit this problem and find
direct correspondence for the four-point conformal block. We thus find an explicit representation for the 
$\widehat{\sll}(2)$ four-point conformal block and hence 
obtain an explicit combinatorial representation for the $\cN=2$ chiral four-point 
conformal block.
\bigskip
\vfill
\end{titlepage}

\tableofcontents

\section{Introduction and Summary}
Conformal invariance and the requirement that the operator algebra be associative reduces the 
evaluation of the correlation functions in any  conformal field theory \cite{BPZ} 
to the calculation of the structure
constants and some universal (model independent) functions known as conformal blocks.
The conformal bootstrap property \cite{Zamolodchikov:1995aa, BBNZ:2007} based on the crossing 
symmetry relations for the four-point correlators
is believed to suffice for the consistency of the conformal field theory. 
The crossing symmetry for the higher multipoint correlation
functions is not expected to give new restrictions on the structure of
the operator algebra. From this standpoint, the four-point conformal 
block plays a key role in CFT.

Conformal blocks are defined as power series of the conformal 
cross-ratios constructed from the space-time coordinates. In principle,
they are completely fixed by the conformal symmetry, although directly
computing their coefficients becomes quite laborious as the power increases. An efficient 
recursive technique for calculating conformal blocks was invented by Al.~B.~Zamolodchikov 
\cite{Zamolodchikov:1987} and further developed in a series of papers (see e.g. \cite{Belavin:2007zz,Hadasz:2007nt,Hadasz:2008dt,Hadasz:2012}). Unfortunately, 
this method essentially depends on the topology of the target space and on the number 
of points in the correlator, and an independent consideration is therefore required 
in each particular case. Recently, following \cite{Alday:2009a}, several relations between
different types of $2d$ conformal field theories and instanton partition functions 
and their associated moduli spaces in $4d$ $\cN\!=\!2$ supersymmetric gauge theories 
have been proposed. This correspondence provides a way to evaluate arbitrary multipoint 
conformal blocks combinatorially. A basic conceptual question in this direction is how to
find the dual instanton moduli space associated with a given CFT symmetry
algebra. Although  there are many examples where the dual pairs have been
found, a general understanding of the correspondence is still lacking,
and further detailed study of the unexplored CFTs seems desirable.

In this context, supersymmetric extensions of the ordinary Virasoro
symmetry are the most natural  objects of investigation. In \cite{Belavin:2011a}, 
the relation between the $\cN\!=\!1$ supersymmetric CFT and $\cN\!=\!2$ $\SU(N)$ gauge 
theories on $\mathbb{R}^4/\mathbb{Z}_2$ was found.  This connection was used 
in \cite{Belavin:2011b} to construct the four-point $\cN\!=\!1$ super Liouville 
conformal  block. Despite the permanent interest in the 
$\cN\!=\!2$ CFT related to its numerous applications in the superstring theory, 
topological field theories, etc., the situation with the conformal blocks in the 
$\cN\!=\!2$ CFT is much less investigated. To our knowledge, essentially only the 
one-point function has been sufficiently well studied 
(see e.g. \cite{Stanishkov:2003, Hosomichi:2004, Eguchi:2004}).
The $\cN\!=\!2$ superconformal field theory was recently considered in the
context of the AGT correspondence \cite{Belavin:2012a}. The relation between
the highest-weight representations of the $\cN\!=\!2$ algebra and the 
so-called relaxed representations of $\widehat{\sll}(2)$ was found by using
the Kazama--Suzuki (KS) coset realization of the $\cN\!=\!2$ superconformal
algebra in terms of the $\widehat{\sll}(2)$ algebra. It was shown that the 
norm of the Whittaker vector, which is related to the Gaiotto limit \cite{Gaiotto:2009b}
of the four-point conformal block, 
is equal to the analytically continued version of the instanton partition function 
for the pure $\cN\!=\!2$ $\SU(2)$  gauge theory with a surface defect. 

In this paper, we study the regular $\cN\!=\!2$ four-point conformal blocks,
focusing on a special class of chiral fields. We show that by means of the
KS map, the $\cN=2$ chiral fields are related  to the ordinary $\widehat{\sll}(2)$ 
primary fields. Since the space of chiral fields is closed under the operator algebra, 
there is no need to consider the effect of the relaxation in this case. Hence, calculating 
the conformal blocks in the chiral ring and calculating the ordinary $\widehat{\sll}(2)$ 
conformal blocks become equivalent. As the previous results \cite{Alday:2010,Kozcaz:2010b, Kanno:2011} 
relating $\widehat{\sll}(2)$ conformal blocks and the instanton partition functions
require special modifications of the conformal blocks, we revisit this correspondence by
using an explicit definition of the dual Laumon moduli space as an orbifold of the 
standard ADHM instanton moduli space. We thus find an explicit representation for the 
$\widehat{\sll}(2)$ four-point conformal block without any modifications and hence 
obtain an explicit combinatorial representation for the $\cN=2$ chiral four-point 
conformal block.

The paper is organized as follows. In section \ref{N2CFT}, we review the $\cN\!=\!2$ 
superconformal algebra and its highest-weight representations and introduce the chiral 
conformal block. In section \ref{SL2}, we review the conformal blocks of the 
$\widehat{\sll}(2)$ algebra and describe the relations between the $\widehat{\sll}(2)$ 
conformal blocks and the instanton partition functions in $\SU(N)$ gauge theories 
with a surface defect. Here, we discuss the problem of the $K$ modification and give 
an explicit representation for the four-point $\widehat{\sll}(2)$ conformal block in
terms of the instanton partition function. Our main result is contained in section \ref{KS}, 
where we use the KS map to find an explicit combinatorial representation for the 
$\cN\!=\!2$ chiral conformal block.

\section{CFT with $\cN=2$ SUSY and conformal blocks}\label{N2CFT}
In the $\cN=2$ superconformal field theory the  holomorphic component of the energy-momentum tensor, 
$T(z)$, is supplemented by two super currents, $G^+(z), G^-(z)$ and by $\U(1)$ current $H(z)$. 
These four generators have conformal weights $2, \frac{3}{2}, \frac{3}{2}, 1$ respectively.

In terms of modes, the $\cN=2$ superalgebra is given by the following commutation relations
\begin{eqnarray}
\label{N2alg}
&&[L_m, L_n] = (m-n) L_{n+m} + \frac{c}{12}(m^3-m)\delta_{m+n,0}\;,
\nonumber\\
&&\left[L_m, H_n\right] = -n H_{m+n}\;,\,\,\,\,\,\,\left[L_m, G_r^{\pm}\right]= 
(\frac{m}{2}-r)G_{m+r}^{\pm}\;,
\nonumber\\
&&\left[H_m, H_n\right] = \frac{c}{3}\; m\; \delta_{m+n,0}\;,\,\, [H_m, G_r^{\pm}] = \pm G_{m+r}^{\pm}\;,
%\,\,
%\{G_r^{\pm}, G_s^{\pm}\} = 0\;,
\\
&&\{G_r^+, G_s^-\} =2 L_{r+s}+(r-s) H_{r+s}+\frac{c}{3}(r^2-\frac 14)\delta_{r+s,0}\;.\nonumber
\end{eqnarray}
Here $c$ is the central charge. In what follows we consider the Neveu--Schwarz sector, where the indices
$m, n$ are integers and $r, s$ are half--integers.
Fields of the $\cN = 2$ CFT belong to the highest weight representations (the Verma modules) of the $\cN=2$ algebra.  
The primary fields $V_{\Delta}^\omega(z)$ correspond to the highest weight vectors for which one has:
\begin{eqnarray}
L_0 V_{\Delta}^\omega=\Delta V_{\Delta}^\omega\,,\,\,\,\,
H_0 V_{\Delta}^\omega=\omega V_{\Delta}^\omega\,,\,\,\,\,
L_{n>0} V_{\Delta}^\omega=0 \,,\,\,\,\,
H_{n>0} V_{\Delta}^\omega=0 \,,\,\,\,\,
G^{\pm}_{r>0} V_{\Delta}^\omega=0\,, 
\end{eqnarray}
where $\Delta$ and $\omega$ are respectively the conformal weight and the $\U(1)$ charge of 
the primary field. The Verma module contains all descendents 
of the given primary field obtained by the action of creating generators on the highest weight 
vector. We introduce special notations for the super--partners of the primary field $V_\Delta^\omega$:
\begin{align}\label{supmultiplet}
&V_{\Delta,\omega}^{+}= G_{-1/2}^{+} V_{\Delta}^{\omega},\quad
V_{\Delta,\omega}^{-}= G_{-1/2}^{-} V_{\Delta}^\omega,\quad \tilde{V}_{\Delta}^{\omega}=\frac 12 \bigg(G_{-1/2}^{+}G_{-1/2}^{-}-G_{-1/2}^{-}G_{-1/2}^{+}\bigg) V_{\Delta}^\omega\,.
\end{align}

Any conformal block is specified by the intermediate channel and defines the holomorphic 
contribution of the descendent states of a given primary one. In the final expression for the 
correlation function, the holomorphic and anti-holomorphic blocks are combined with the structure 
constants and summed over intermediate primary fields allowed by the fusion rules. Our concrete calculations
focus on the conformal blocks of four primary fields $V_{\Delta_1}^{\omega_1}, V_{\Delta_2}^{\omega_2}, V_{\Delta_3}^{\omega_3}, V_{\Delta_4}^{\omega_4}$ with the primary field
$V_{\Delta}^{\omega}$ in the intermediate channel such that $\omega_1+\omega_2=\omega=\omega_3+\omega_4$. Inserting a complete set of states in the correlation function one finds the following schematic expression 
\begin{equation}
\mathcal{F}_{{\scriptscriptstyle \cN_2 }}\left(\left.\begin{array}[c]{cc}%
 \Delta_1,\omega_1&\Delta_4,\omega_4\\\Delta_2,\omega_2&\Delta_3,\omega_3\end{array}\right|\Delta,\omega\,\bigg|\,z\right)=
 \sum^{\infty}_{l=0} z^l\sum_{\alpha,\beta} \langle V_{\Delta_1}^{\omega_1} | V_{\Delta_2}^{\omega_2} | \alpha \rangle_{l} \times  
 ( K^{-1}_{\Delta,\omega}(l))_{\alpha\beta} \times {}_{l}\langle \beta | V_{\Delta_3}^{\omega_3} | V_{\Delta_4}^{\omega_4}\rangle\,,
\end{equation}
where the sum is performed over all $l$th level descendants $|\alpha\rangle_{l}$ of the field $V_{\Delta}^{\omega}$ and $H_0|\alpha\rangle_{l}=\omega|\alpha\rangle_{l}$. This restriction arises because of the conservation of the $\U(1)$ current.
The matrix $K^{-1}_{\Delta,\omega} (l)$ is the inverse
of the corresponding block of Gram/Shapovalov matrix   
(($K_{\Delta,\omega} (l))_{\alpha\beta}= {}_{l}\langle \alpha|\beta\rangle_{l}$). 
The triple vertices 
$\langle V_{\Delta_1}^{\omega_1}| V_{\Delta_2}^{\omega_2} | \alpha \rangle_{l}$ and ${}_{l}\langle \beta | V_{\Delta_3}^{\omega_3} | V_{\Delta_4}^{\omega_4}\rangle$
can be obtained by using the following commutation relations for the vertex operator 
$V(z)\doteqdot V_{\Delta}^{\omega}(z)$ and its superpartners
\begin{eqnarray}\label{commV}
&&[L_k, V(z)]=((k+1)\Delta z^k +z^{k+1}\partial) V(z)\,, \nonumber\\
&&[L_k, V^{\pm}(z)]=((k+1)(\Delta+\frac 12) z^{k} +z^{k+1}\partial) V^{\pm}(z)\,,\nonumber\\
&&[L_k, \tilde{V}(z)]=k(k+1)\frac{\omega}{2} z^{k-1} V(z)+\big((k+1)(\Delta+1)z^k+z^{k+1}\partial\big)\tilde{V}(z)\,,\nonumber\\
&&[H_k, V(z)]=\omega z^k V(z)\,,\nonumber\\
&&[H_k, V^{\pm}(z)]=(\omega\pm 1) z^k V^{\pm}(z)\,,\\
&&[H_k, \tilde{V}(z)]=2 k\Delta  z^{k-1} V(z)+\omega z^k \tilde{V}(z)\,,\nonumber\\
&&[G_k^{\pm}, V(z)]=z^{k+\frac 12} V^{\pm}(z)\,, \nonumber\\
&&\{G_k^{\pm}, V^{\pm}(z)\}=0\,,\nonumber\\
&&\{G_k^{\pm}, V^{\mp}(z)\}=((k+\frac 12)(2\Delta\pm \omega) z^{k-\frac 12} +z^{k+\frac 12}\partial) V(z)
\pm z^{k+\frac 12} \tilde{V}(z)\,,\nonumber\\
&&[G_k^{\pm}, \tilde{V}(z)]=\mp(k+\frac 12)(2\Delta\pm\omega+1) z^{k-\frac 12} V^{\pm}(z)\mp z^{k+\frac 12}\partial V^{\pm}(z)\,,\nonumber
\end{eqnarray}
and taking into account that 
$\langle V_{\Delta_1}^{\omega_1}| V_{\Delta_2}^{\omega_2} (z)| V_{\Delta}^{\omega} \rangle
\propto z^{\Delta-\Delta_2-\Delta_1}$. The normalization is fixed by the requirement 
that the first coefficient in $z$-expansion of the conformal block is one. 

In what follows we are interested in the chiral primary fields \cite{Vafa:1989, Mussardo:1989}. The primary field is called chiral if it satisfies  
\begin{equation}\label{Chiral}
G^+_{-1/2}V_{\Delta}^{\omega}=0\,.
\end{equation}
That is half of its super-partners vanish. The condition (\ref{Chiral}) implies $\omega=2\Delta$, 
meaning that a chiral primary field depends on one continuous parameter instead of two.  
Since below we are always dealing with $\cN=2$ chiral fields, we denote them simply by 
$V_{\Delta}$(z). The commutation relations \eqref{commV} for the chiral states are reduced to the 
following system 
  \begin{eqnarray}\label{chainBPS}
  &&[L_k, V(z)]=((k+1)\Delta z^k +z^{k+1}\partial) V(z)\,, \nonumber\\
&&[L_k, V^{-}(z)]=((k+1)(\Delta+\frac 12) z^{k} +z^{k+1}\partial) V^{-}(z)\,,\nonumber\\
&&[H_k, V(z)]=\omega z^k V(z)\,,\nonumber\\
&&[H_k, V^{-}(z)]=(\omega- 1) z^k V^{-}(z)\,,\\
&&[G_k^{+}, V(z)]=0 \,,\nonumber\\
&&[G_k^{-}, V(z)]=z^{k+\frac 12} V^{-}(z) \,,\nonumber\\
&&\{G_k^{+}, V^{-}(z)\}=((k+\frac 12)(2\Delta+ \omega) z^{k-\frac 12} +z^{k+\frac 12}\partial) V(z)\,,\nonumber\\
&&\{G_k^{-}, V^{-}(z)\}=0\,.\nonumber
\end{eqnarray}
The conservation of the $\U(1)$ current together with the condition \eqref{Chiral} reduce
the number of the parameters of the conformal blocks. For the four-point conformal block 
we choose free independent parameters to be $\Delta_2, \Delta_3$ and $\Delta$. 
The coefficients in the power expansion 
\begin{eqnarray}
\mathcal{F}_{{\scriptscriptstyle \cN_2 }}\!\!\left(\!\!\left.\begin{array}[c]{cc}%
 \Delta-\Delta_2&\Delta-\Delta_3\\\Delta_2&\Delta_3\end{array}\right|\Delta\,\bigg|\,z\right)
 \!\equiv\! \mathcal{F}_{{\scriptscriptstyle \cN_2 }} (\Delta_2,\Delta_3,\Delta;z)
\! =
\!\sum_{l=0}^{\infty} z^l  \mathcal{F}_{{\scriptscriptstyle \cN_2}}^{(l)} (\Delta_2,\Delta_3,\Delta),
\end{eqnarray}
can be found order by order from \eqref{chainBPS}. The explicit formulae for the first 
three coefficients are listed below
\begin{eqnarray}
&&\mathcal{F}_{{\scriptscriptstyle \cN_2}}^{(0)} (\Delta_2,\Delta_3,\Delta)= 1\nonumber\,,\\
&&\mathcal{F}_{{\scriptscriptstyle \cN_2}}^{(1)} (\Delta_2,\Delta_3,\Delta)= \frac{2\Delta_2\Delta_3}{\Delta}\nonumber\,,\\
&&\mathcal{F}_{{\scriptscriptstyle \cN_2}}^{(2)} (\Delta_2,\Delta_3,\Delta)=\Delta_2 \Delta_3\times \nonumber\\
&&\frac{ 
\big(12 \Delta_2 \Delta_3 - 2 (\Delta_2 + \Delta_3) + \hat c (1 + 2 \Delta_2) (1 + 2 \Delta_3) + 
   4\Delta (1 - 2 \Delta_2) (1 - 2 \Delta_3) + 16 \Delta^2-1\big)}{\Delta (1 + 2 \Delta) (4 \Delta + \hat c - 1)}
\nonumber\,,
\end{eqnarray}
where $\hat c\equiv c/3$.

\section{Affine $sl(2)$ blocks and instanton partition functions}\label{SL2}
In \cite{Alday:2010} the correspondence between $\widehat{\sll}(2)$ conformal field theory
and four-dimensional $\cN=2$ gauge field theories was found. This correspondence provides
explicit representation for $\widehat{\sll}(2)$ conformal blocks with insertion of a certain
$K$ operator in terms of Nekrasov instanton partition functions on the Laumon moduli spaces.
For computing $\cN=2$ conformal blocks in the next section, the proper $\widehat{\sll}(2)$  
conformal block is relevant. It was noticed in \cite{KIns2010a,KIns2010b,KIns2010c,KIns2010d,KIns2010e} 
that for four-point conformal
block the effect of the insertion of the $K$ operator can be replaced by considering a modified
mapping between the gauge theory and the CFT variables. In this section we briefly recall this 
correspondence and describe the necessary modification in case of four-point function.

The  $\widehat{\sll}(2)$ commutation relations are 
\begin{eqnarray}\label{sl2algebra}
[J_n^0,J_m^0] = \frac{k}{2} \,n\, \delta_{n+m,0} \,,\quad 
[J_n^0,J_m^{\pm}] = \pm J_{n+m}^{\pm}  \,, \quad
[J_n^+,J_m^-] = 2J_{n+m}^0 + k\,  n\, \delta_{n+m,0} \,,
\end{eqnarray}
where $n, m\in\ZZ$ and $k$ is the level of  $\widehat{\sll}(2)$. 
The highest-weight representation (module) is defined by imposing  the following
requirements on the highest-weight state $|j\rangle$:
\begin{equation}\label{standHWV}
J_0^0 |j\rangle = j  |j\rangle\,,\qquad J_{n>0}^0 |j\rangle =J_{n>0}^- |j\rangle =J_{n\ge0}^+ |j\rangle =0\,,
\end{equation}
and is freely generated by the action of the remaining modes $J_{n}^A$.
The energy level (minus the sum of the mode numbers) and $J_0^0$ charge define two natural gradings 
in the $\widehat{\sll}(2)$ module.
Similar to the $N = 2$ case the matrix of inner products of descendants  has a block-diagonal
structure with respect to these gradings.

We are interested in the conformal blocks of the primary $\widehat{\sll}(2)$ fields $V_{j_i}(z_i)$. 
In four-point case one finds
\begin{equation}
\mathcal{F}^{(n)}_{{\scriptscriptstyle \widehat{\sll}_2 }}\left(\left.\begin{array}[c]{cc}%
 j_1&j_4\\j_2&j_3\end{array}\right|j\,\bigg|\,z\right)=
 \sum^{\infty}_{l=0} z^l\sum_{\alpha,\beta} \langle V_{j_1} | V_{j_2} | \alpha \rangle_{l} \times  
 ( K^{-1}_{j,n}(l))_{\alpha \beta} \times {}_{l}\langle \beta | V_{j_3} | V_{j_4}\rangle\,,
\end{equation}
where $|\alpha\rangle_{l}$,  $|\beta\rangle_{l}$ are $l$th level descendants of the field $V_j$ such that
$J_0^0|\alpha\rangle_{l}=(j+n)|\alpha\rangle_{l}$ and $n$ is some integer $n\geq -l$.
The matrix $K^{-1}_{j,n} (l)$ is the inverse
of the corresponding block of Gram/Shapovalov matrix. 
The triple vertices 
$\langle V_{j_1}| V_{j_2} | \alpha \rangle_{l}$ and ${}_{l}\langle \beta | V_{j_3} | V_{j_4}\rangle$
are obtained by introducing auxiliary variable $x$ on which $sl(2)$ acts and
using the following commutation relations for the vertex operator 
$V_j(z,x)$ 
\begin{equation}
[J_n^a,V_j(z,x)]=z^n \mathcal{D}^a V_j(x,z)\,,
\end{equation}
where the differential operators $\mathcal{D}^a$ are given by
\begin{eqnarray}
&&\mathcal{D}^+=-x^2\partial_x+2 j x\,,\nonumber\\
&&\mathcal{D}^0=-x\partial_x+ j \,,\nonumber\\
&&\mathcal{D}^-=-\partial_x\,.
\end{eqnarray}
and taking into account that 
$\langle V_{j_1}| V_{j_2}(x,z)| V_{j} \rangle \propto x^{j_1+j_2-j}$.
One finds the following first coefficients of the series expansion
\begin{align}
&\sum_{n} x^n \mathcal{F}^{(n)}_{{\scriptscriptstyle \widehat{\sll}_2 }}\left(\left.\begin{array}[c]{cc}%
 j_1&j_4\\j_2&j_3\end{array}\right|j\,\bigg|\,z\right)=\nonumber\\
& =z^0 x^0 + \frac{(j_1 - j_2 - j) (j_3 - j_4 + j)}{2 j} z^0 x^1 +
  \frac{(j_3 + j_4 - j) (j_1 + j_2 - j)}{2 j - k} z^1 x^{-1} +\nonumber\\
& +\bigg(\frac{(j_4 - j) (j_1 + j_1^2 - j_2 - j_2^2 - j - j^2)
-(j_1 - j) (j_3 + j_4 - j) (1 + j_3 - j_4 + j)}{j (2 + k)}+\\
& +\frac{(j_1 + j_2 - j) (1 - j_1 + j_2 + j) (j_3 + j_4 - j) (1 + j_3 - j_4 + j) k}{2 j (2 + k) (-2 j + k)}\bigg) 
z^1 x^0 +\cdots \nonumber
\end{align}
From \eqref{standHWV} we note that the power of $x$ in the denominator cannot be lager then the power of $z$.

The dual AGT description of $\widehat{\sll}(N)$ conformal blocks involves the instanton partition functions of $\cN\!=\!2$ $\SU(N)$ gauge theories with a certain surface defect \cite{Alday:2010,Kozcaz:2010b, Kanno:2011}. The relevant  instanton moduli space is known as Laumon space \cite{Braverman:2004a, Finkelberg:2010, Feigin:2008}.  According to the general rule, to consider the four-point conformal block in addition to the gauge field multiplet,
four matter fields in the fundamental representation should be taken into account in the instanton calculations. 
Below we formulate the result for this particular case. Because of the Nekrasov's deformation 
the functional integral of the gauge theory is localized  and the evaluation of the instanton partition function is reduced to the calculation of the bosonic and fermionic determinants in the fixed points of some vector field 
$\vec v$ acting on the Laumon space. The Laumon space itself 
is defined by imposing some additional $Z_2$ symmetry restriction on the ordinary ADHM moduli space, and
hence can be considered as its $Z_2$ symmetric sub-space \footnote{Our approach follows \cite{Belavin:2012a} where the precise definitions can be found.}.
This manifold contains a number of disconnected components,
characterized  by two topological numbers $k_1$ and $k_2$. For each component, these integers give the dimensions of its subspaces having different parities with respect to the $Z_2$ symmetry. The fixed points are labelled by pairs of Young diagrams $\vec Y=(Y_1,Y_2)$. 
The form of the Young diagrams can be translated into the corresponding solution of the ADHM relations.
The parity characteristic $0,1$ is assigned to each box in the Young diagrams related to the fixed point. 
In our case, $(1,1)$ boxes in the first and in the second diagrams have $q_1=0$ (white) and $q_2=1$ (black) respectively.
The parity of the boxes is the same in each column  and differ between 
two neighbour columns of the diagram. Each fixed point belongs to one or another connected components of the Laumon 
moduli space. The corresponding instanton numbers $k_1,k_2$ are given by
\begin{equation} \label{Yconstr}
\sum_{j=1}  [ (Y_1)_{2j-1} + (Y_2)_{2j} ] = k_1\,, \qquad  \sum_{j=1} [ (Y_1)_{2j} + (Y_2)_{2j-1} ] = k_2 \,,
\end{equation}
where $(Y_\alpha)_j$ denotes the height of column $j$ in the diagram $Y_\alpha$. In other words,
$k_1$ and $k_2$ are the total numbers of white and black boxes in $(Y_1,Y_2)$. 

The instanton partition function  is given by  
\begin{equation}\label{Z22}
\mathcal{Z}_{\text{inst}}(m_i,\vec a|y_1,y_2)  = \sum_{\vec Y} Z_{\vec Y}(m_i,\vec a)  y_1^{k_1} y_2^{k_2} \,,
\end{equation}
where $\vec a=(a,-a)$ is the vacuum expectation value of the gauge field and $m_i$ are the masses of the hypers.
In each fixed point $Z_{\vec Y}(m_i,\vec a)$ gets contributions from the gauge vector multiplet and from the (anti-)fundamental matter fields 
\begin{equation}
Z_{\vec Y}(m_i,\vec a)=Z_{\vec Y}^{\text{vec}}(\vec a)Z_{\vec Y}^{\text{\bf f}}(m_i,\vec a)\,.
\end{equation}
To evaluate the determinant coming from the gauge multiplet in the fixed point one needs to know the eigenvalues of
the vector field $\vec v$ on the tangent space. The problem is reduced
to the evaluation of the matrix elements between basis vectors represented by the cells of the Young diagrams
and leads to the standard form of the vector contribution \cite{Alday:2009a}. In
case of the Laumon  space the tangent space is reduced by the additional requirement of $Z_2$ symmetry,
hence the eigenvectors out of the new tangent space should be excluded, this gives \cite{Belavin:2012a}: 
\begin{equation}\label{det-vp-Zp}
 Z_{\vec Y}^{\text{vec}}(\vec a) \!=\!\prod_{\alpha,\beta=1}^{2} 
 \prod_{s_\alpha \in \scriptscriptstyle{Y_{\alpha,\beta}^{(0)}}} \!\!\!
    E^{-1}_{\scriptscriptstyle{Y_{\alpha}},\scriptscriptstyle{Y_{\beta}}}(a_{\beta}{-}a_{\alpha}|s_\alpha )\!\!\!
 \prod_{s_\alpha \in \scriptscriptstyle{Y_{\alpha,\beta}^{(1)}}}  \!\!
   \bigl(\epsilon{-}E_{\scriptscriptstyle{Y_{\alpha}},\scriptscriptstyle{Y_{\beta}}}(a_{\beta}{-}a_{\alpha}|s_\alpha )\bigr)^{-1},
\end{equation}
where $\epsilon=\epsilon_1+\epsilon_2$ ($\epsilon_{1,2}$ are Omega-background deformation parameters) and  
\begin{equation}\label{E-def}
    E_{\scriptscriptstyle{Y_{\alpha}},\scriptscriptstyle{Y_{\beta}}}(x|s_\alpha )=x+\epsilon_{1}(L_{\scriptscriptstyle{Y_{\alpha}}}(s_\alpha )+1)-\epsilon_{2}\,A_{\scriptscriptstyle{Y_\beta}}(s_\alpha)\,.
\end{equation}
The leg-factor $L_{\scriptscriptstyle{Y_{\beta}}}(s_\alpha )$ and the arm-factor  $A_{\scriptscriptstyle{Y_\beta}}(s_\alpha )$
are correspondingly the lengths from the box $s_\alpha \in Y_\alpha$ to the end of the column and to
the end of the row with respect to the 
Young diagram $Y_\beta$. 
The regions $ Y_{\alpha,\beta}^{(g)}$ ($g=0,1$) in \eqref{det-vp-Zp} are defined as 
\begin{equation}\label{regions}
Y_{\alpha,\beta}^{(g)}=\{s_\alpha \in Y_{\alpha}\, |\, q_\beta-q_\alpha-A_\beta(s_\alpha )=g \!\! \mod 2 \}\,.
\end{equation}
Because of the fermionic zero-modes, hypermultiplets give additional contributions  in the instanton 
partition function. This effect can be described by attaching additional fiber to the fixed point. 
The corresponding contribution of the fundamental hypermultiplets
with masses $m_i$ evaluated on the ADHM moduli space  looks as \cite{Alday:2009a}
\begin{equation}
Z_{\vec Y}^{\text{{\bf f}\,(0)}}(m_i,\vec{a})=\prod_{i=1}^{4}\prod_{\alpha=1}^{2}
    \prod_{s\in Y_{\alpha}}\bigl(\phi(a_{\alpha},s)+m_i\bigr),
\end{equation}
where
\begin{equation}
\phi(x, s) =x+(i_s-1)\epsilon_2+(j_s-1)\epsilon_1\,,
\end{equation}
where in our conventions $s$ box is found in the $i$th column and in the $j$th row of the diagram.
In the Laumon space we impose some  restrictions on the set of eigenvectors for the fundamental multiplets
of the same origin as in the vector case.
This gives the following form of the contribution of the fundamental hyper multiplets
\begin{eqnarray}\label{Zf}
Z_{\vec Y}^{\text{\bf f}}(m_i,\vec{a})=(-1)^{|\vec Y|} &&
  \!\!\! \!\!\! \!\!\!   
  \!\!\!  \! \prod_{\substack{s\in \scriptscriptstyle{Y_{1}}\\ s-\text{white}}}  \!\!\!
   \!\!   \bigl(\phi(a,s)+m_1\bigr)\bigl(\phi(a,s)+m_4\bigr) 
  \!\!   \!\!    \!
      \prod_{\substack{s\in \scriptscriptstyle{Y_{1}}\\ s-\text{black}}}  \!\!\!  \!\! 
 \!\!    \bigl(\phi(a,s)+m_2\bigr)\bigl(\phi(a,s)+m_3\bigr) \\
  \times &&   \!\!\!  \!\!\! \!\!\! \!
  \!\!\!   \prod_{\substack{s\in \scriptscriptstyle{Y_{2}}\\ s-\text{white}}}  \!\!\!  \!\! 
   \bigl(\phi(-a,s)+m_1\bigr)\bigl(\phi(-a,s)+m_4\bigr)
 \!\!\!    \!\!   \prod_{\substack{s\in \scriptscriptstyle{Y_{2}}\\ s-\text{black}}}  \!\!\!  \!\! 
   \bigl(\phi(-a,s)+m_2\bigr)\bigl(\phi(-a,s)+m_3\bigr). \nonumber
\end{eqnarray}
This result can be alternatively derived from the general expression for the 
bifundamental multiplet contribution which can be found in \cite{Alday:2010}.  
We note that to make our results compatible with those in  \cite{Alday:2010,Kozcaz:2010b} 
the following redefinition of the parameters is required
\begin{equation}
\epsilon_2 \rightarrow \frac{\epsilon_2}{2}\,,\quad a \rightarrow a + \frac{\epsilon_2}{4}\,,
\end{equation}
and
\begin{equation}
 m_1 \rightarrow m_1 + \epsilon_1+\frac{3\epsilon_2}{4}\,,\quad 
 m_2 \rightarrow m_2 + \epsilon_1+\frac{\epsilon_2}{4}\,, \quad 
 m_3 \rightarrow m_3 + \frac{\epsilon_2}{4}\,, \quad
 m_4 \rightarrow m_4 - \frac{\epsilon_2}{4}\,.
\end{equation}

In \cite{Alday:2010,Kozcaz:2010b} the relation between the instanton partition functions and 
$\widehat{\sll}(2)$ conformal blocks modified by the insertions of some additional operator was found. 
For our purposes we will need to have similar relation for the conformal block without any modification. 
For four-point function this relation can be obtained provided  
%\begin{align} \label{AGTold}
%&j_1 = -\frac{\epsilon_1+\epsilon_2 + m_1 - m_2}{2 \epsilon_1}, 
%j_2 = -\frac{2\epsilon_1 + \epsilon_2 + m_1 + m_2}{2 \epsilon_1}, 
%j_3 = -\frac{m_3 + m_4}{2 \epsilon_1}, j_4 = -\frac{\epsilon_1 - m_3 + m_4}{2 \epsilon_1}, \nonumber\\
%&\qquad\qquad\qquad j = -\frac 12 + \frac{a}{\epsilon_1}\,,\qquad k = -2 - \frac{\epsilon_2}{\epsilon_1}\,, 
%\qquad y_1=-x \,, \qquad y_2 = -\frac{z}{x}\,.
%\end{align}
\begin{align} \label{AGT}
&j_1 = -\frac{\epsilon_1 + \epsilon_2 + m_1 - m_2}{2 \epsilon_1},\,\, j_2 = -\frac{m_1 + m_2}{2 \epsilon_1}, \,\,
 j_3 = -\frac{m_3 + m_4}{2 \epsilon_1},\,\, j_4 = -\frac{\epsilon_1 + \epsilon_2 - m_3 + m_4}{2 \epsilon_1}, \nonumber\\
&\qquad\qquad j = -\frac{\epsilon_1 + \epsilon_2 - 2 a}{2 \epsilon_1}\,,\qquad  
k = -\frac{2 (\epsilon_1 + \epsilon_2)}{\epsilon_1}\,, \qquad y_1=-x \,, \qquad y_2 = -\frac{z}{x}\,.
\end{align}
Using this map one finds the following connection between the instanton partition function
of the surface operator given in \eqref{Z22}-\eqref{Zf} and the four-point $\widehat{\sll}(2)$ conformal block function
\begin{equation}\label{sl2inst}
\mathcal{Z}_{\text{inst}}(m_i,\vec a|y_1,y_2)=(1-z)^{-\frac{2j_2j_3}{k+2}}\sum_n x^n\mathcal{F}^{(n)}_{{\scriptscriptstyle \widehat{\sll}_2 }}\left(\left.\begin{array}[c]{cc}%
 j_1&j_4\\j_2&j_3\end{array}\right|j\,\bigg|\,z\right) \,.
\end{equation}
We have checked this relation up to the fifth order. 

\section{KS mapping and $\cN=2$ chiral conformal blocks}\label{KS}
The $\cN\!=\!2$ superconformal algebra can be realised in terms of $\widehat{\sll}(2)$ and a free complex fermion $\psi(z)$ ($\bar\psi(z)$) by using Kazama--Suzuki coset construction~\cite{DiVecchia:1986, Kazama:1988}
\begin{equation} \label{21coset}
\frac{\hat{\sll}(2)_k\! \times\widehat{\ul}(1)}{\widehat{\ul}(1)} \,,
\end{equation}
where $k$ is the level of the affine $\widehat{\sll}(2)$ algebra. 
The $\widehat{\ul}(1)$ algebra in the numerator of the coset (\ref{21coset}) corresponds to the
conserved current $\psi \bar \psi(z)$ while the $\widehat{\ul}(1)$ sub-algebra in the denominator 
is generated by 
\begin{equation}  \label{Kdef}
K(z)=J^0(z)-\psi\bar\psi(z)\,.
\end{equation} 
The $\cN\!=\!2$ superconformal algebra
is defined as sub-algebra in the numerator of \eqref{21coset} which commutes with $K(z)$. This requirement
fixes 
\begin{equation}
H(z)=\frac{1}{2(k+2)}J^0(z)+\frac{k}{k+2}\psi\bar\psi(z).
\end{equation}
The odd generators $G^{+}$ and $G^{-}$ are produced as
\begin{equation}  \label{QGsl}
  G^+(z)= \sqrt{\frac{2}{k{+}2}}\, \psi(z) J^+(z)\,,\qquad 
  G^-(z)= \sqrt{\frac{2}{k{+}2}}\, \bar \psi(z) J^-(z)\,,
\end{equation}
By construction they have dimensions $3/2$ and vanishing OPEs with $K(z)$. 
The stress-energy tensor $T(z)$ of the $\cN=2$ algebra is uniquely fixed 
from the $G^+(z)G^-(w)$ operator product expansion.
The central charge of the $\cN\!=\!2$ algebra is expressed in terms of the $\widehat{\sll}(2)$ level $k$ as 
\begin{equation}\label{ck}
c={\frac{3k}{k+2}}\,.
\end{equation}

To build up the highest weight representation, we need to specify the primary fields 
of the $\cN=2$ algebra. General primary fields can be constructed from the 
relaxed $\widehat{\sll}(2)$ primary fields $\Phi_j^{\lambda}(z)$ \cite{Feigin:1997} dressed by the exponential
fields $e^{\beta \phi}$, where the bosonic field $\phi$ is related to the current $K(z)$ as
\begin{equation}
K(z)=\partial \phi(z)\,.
\end{equation} 
The parameter $\beta$ is again fixed by the requirement of having vanishing OPEs between $K(z)$
and the $\cN\!=\!2$ primary fields
\begin{equation}
V^{\omega}_{\Delta}(z)=\Phi_j^{\lambda}(z) e^{\beta\phi}(z)\,.
\end{equation}
This requirement fixes also the relations between the parameters  
\begin{equation} \label{dj}
\Delta=\frac{j(j{+}1)-\lambda^2}{k+2}\,,\qquad  \omega=\frac{2\lambda}{k+2}\,.
\end{equation}

For the $\cN=2$ chiral primary fields one finds that $\lambda=j$, hence the 
corresponding  $\widehat{\sll}(2)$ filed is the highest weight vector $\Phi_j(z)$ 
satisfying \eqref{standHWV} and the mapping between parameters is
\begin{equation} \label{djchiral}
\omega=2\Delta=\frac{2j}{k+2}.
\end{equation}

Because $\widehat{\sll}(2)$ and the free fermion sectors do not interact, the 
$\cN=2$ conformal blocks are given by the products of the conformal blocks of the
free exponential fields and the conformal blocks of the primary $\widehat{\sll}(2)$ 
fields. The former can be computed explicitly, so that in four-point 
case, one finds the following relation
\begin{equation}\label{N2sl2}
\mathcal{F}_{{\scriptscriptstyle \cN_2 }}\!\left(\begin{array}[c]{cc}%
 \Delta-\Delta_2&\Delta-\Delta_3\\\Delta_2&\Delta_3\end{array}\bigg|\Delta\bigg|z\right)=
 (1-z)^{\frac{2j_2j_3}{k+2}}
 \mathcal{F}^{(0)}_{{\scriptscriptstyle \widehat{\sll}_2 }}\left(\begin{array}[c]{cc}%
 j-j_2&j-j_3\\j_2&j_3\end{array}\bigg|j\bigg|z\right)\!.
\end{equation}
Taking the results of the previous section into account, one finds that $\cN=2$ chiral conformal
block can be expressed in terms of the Nekrasov's instanton partition function.
Combining \eqref{sl2inst} and \eqref{N2sl2} we obtain the following explicit combinatorial 
representation for the $\cN\!=\!2$ chiral four-point 
block (with $k_{1,2}$ being defined in \eqref{Yconstr}):
 \begin{align}\label{N2block1}
\mathcal{F}_{{\scriptscriptstyle \cN_2 }} (\Delta_2,\Delta_3,\Delta;z)
=(1-z)^{\frac{8\Delta_2\Delta_3}{1-\hat c}}\sum_{\substack{Y_1,Y_2\\k_1=k_2}} 
\frac{K_{\text{f}}(Y_1,Y_2)}{K_{\text{v}}(Y_1,Y_2)} z^{k_2}\,,
 \end{align}
where
 \begin{eqnarray}
\label{KY1Y2vec}
  \!\!\!  \!\!\!  \!\!\! K_{\text{v}}(Y_1,Y_2) =  &&
  \!\!\! \!\!\! \!\!\!   
  \!\!\!   \prod_{\substack{s\in \scriptscriptstyle{Y_{1}}\\ A(s_1)-\text{odd}}}  \!\!\!
   \!\!   E_{\scriptscriptstyle{Y_1},\scriptscriptstyle{Y_2}}(2 a |s)
   (\epsilon- E_{\scriptscriptstyle{Y_{1}},\scriptscriptstyle{Y_{1}}}(0|s))
  \!\!   \!\!    \!\!\!     \prod_{\substack{s\in \scriptscriptstyle{Y_{1}}\\ A(s)-\text{even}}}  \!\!\!  \!\! 
 \!\!   E_{\scriptscriptstyle{Y_{1}},\scriptscriptstyle{Y_{1}}}(0|s)      
        \bigl(\epsilon-E_{\scriptscriptstyle{Y_{1}},\scriptscriptstyle{Y_{2}}}(2a|s)\bigr) \,\,\,\\
   &&   \!\!\!  \!\!\! \!\!\! \!\!  \times  \!
  \!\!\!  \!\! \prod_{\substack{s\in \scriptscriptstyle{Y_{2}}\\ A(s)-\text{odd}}}  \!\!\!  \!\! 
   E_{\scriptscriptstyle{Y_2},\scriptscriptstyle{Y_1}}(-2a|s)(\epsilon-E_{\scriptscriptstyle{Y_{2}},\scriptscriptstyle{Y_{2}}}(0|s))
 \!\!\!    \!\!  \!\!\!   \prod_{\substack{s\in \scriptscriptstyle{Y_{2}}\\ A(s)-\text{even}}}  \!\!\!  \!\! 
    E{\scriptscriptstyle{Y_{2}},\scriptscriptstyle{Y_{2}}}(0|s)\bigl(\epsilon-E_{\scriptscriptstyle{Y_{2}},\scriptscriptstyle{Y_{1}}}(-2a|s\bigr), \nonumber
\end{eqnarray}
and
\begin{eqnarray}
\label{KY1Y2fun}
   \!\!\! K_{\text{f}}(Y_1,Y_2) = &&
  \!\!\! \!\!\! \!\!\!   
  \!\!\!   \prod_{\substack{s\in \scriptscriptstyle{Y_{1}}\\ s-\text{white}}}  \!\!\!
   \!\!   \bigl(\phi(a,s)+m_1\bigr)\bigl(\phi(a,s)+m_4\bigr) 
  \!\!   \!\!    \!\! 
      \prod_{\substack{s\in \scriptscriptstyle{Y_{1}}\\ s-\text{black}}}  \!\!\!  \!\! 
 \!\!    \bigl(\phi(a,s)+m_2\bigr)\bigl(\phi(a,s)+m_3\bigr) \\
  \times &&   \!\!  \!\!\! \!\!\! \!
  \!\!\!  \!\! \prod_{\substack{s\in \scriptscriptstyle{Y_{2}}\\ s-\text{white}}}  \!\!\!  \!\! 
   \bigl(\phi(-a,s)+m_1\bigr)\bigl(\phi(-a,s)+m_4\bigr)
 \!\!\!    \!\!  \! \prod_{\substack{s\in \scriptscriptstyle{Y_{2}}\\ s-\text{black}}}  \!\!\!  \!\! 
   \bigl(\phi(-a,s)+m_2\bigr)\bigl(\phi(-a,s)+m_3\bigr). \nonumber
\end{eqnarray}
From \eqref{AGT} it follows that $m_1=m_4=-a$. Taking into account this constraint, one finds that if
the first diagram is not empty,  the corresponding contribution $K_{\text{f}}(Y_1,Y_2)$
is zero. The sum in \eqref{N2block1} is reduced to the sum over one Young diagram  
 \begin{eqnarray}\label{N2block2}
&&\mathcal{F}_{{\scriptscriptstyle \cN_2 }} (\Delta_2,\Delta_3,\Delta;z)
=(1-z)^{\frac{8\Delta_2\Delta_3}{1-\hat c}}\times\\
&&\times\!\!\!\sum_{\substack{Y\\k_1=k_2}} 
\frac{ \prod_{\substack{s\in \scriptscriptstyle{Y}\\ s-\text{white}}}  \!\!\!  \!\! 
   \phi(-2a,s)\phi(-2a,s)
  \prod_{\substack{s\in \scriptscriptstyle{Y}\\ s-\text{black}}}
  \bigl(\phi(-a,s)+m_2\bigr)\bigl(\phi(-a,s)+m_3\bigr)}
 {\prod_{\substack{s\in \scriptscriptstyle{Y}\\ A(s)-\text{odd}}}  
   \!\!\! \!\! E_{\scriptscriptstyle{Y},\varnothing}(-2a|s)(\epsilon-E_{\scriptscriptstyle{Y},\scriptscriptstyle{Y}}(0|s))  \prod_{\substack{s\in \scriptscriptstyle{Y}\\ A(s)-\text{even}}}  \!\!\!  \!\! 
    E{\scriptscriptstyle{Y},\scriptscriptstyle{Y}}(0|s)\bigl(\epsilon-E_{\scriptscriptstyle{Y},\varnothing}(-2a|s\bigr)} z^{k_2}\,,\nonumber
 \end{eqnarray}
where the direct mapping between $\cN=2$ CFT and gauge theory parameters is
\begin{align} \label{N2SU2map}
\Delta_2 = \frac{m_2 - a}{4 \epsilon_2}\,, \quad
\Delta_3 = \frac{m_3 - a}{4 \epsilon_2}\,,\quad 
\Delta = \frac{\epsilon_1 + \epsilon_2 - 2 a}{4 \epsilon_2}\,,\quad 
\hat c = 1+\frac{\epsilon_1}{\epsilon_2}\,.
\end{align}

Hence we come to our main result, the following closed expression for the  conformal block function of 
four $\cN=2$ chiral primary operators:
 \begin{eqnarray}\label{N2block3}
&&\mathcal{F}_{{\scriptscriptstyle \cN_2 }} (\Delta_2,\Delta_3,\Delta;z)
=(1-z)^{\frac{8\Delta_2\Delta_3}{1-\hat c}}\sum_{\substack{Y\\k_1=k_2}}
\mathcal{B}_Y(\Delta_2,\Delta_3,\Delta) z^{k_2}\,,
\end{eqnarray}
with the coefficients 
\begin{eqnarray}\label{N2coef}
&&\mathcal{B}_Y(\Delta_2,\Delta_3,\Delta)= \\
&&\frac{\prod_{\substack{(i,j)\in \scriptscriptstyle{Y}\\ i-\text{even}}}    
   \bigl(4\Delta-\hat c+(i-1)+(j-1)(\hat c-1)\bigr)^2}
 {    \prod_{\substack{(i,j)\in \scriptscriptstyle{Y}\\ i-\text{even}}}  \bigl(\hat c-4\Delta-(i-1)-(Y_i'-j)(\hat c-1)\bigr) \prod_{\!\!\!\!\!\!\substack{(i,j)\in \scriptscriptstyle{Y}\\ (Y_j-i)-\text{even}}} \!\!\!  \!\! 
   \bigl((i-Y_j)+(Y_i'-j+1)(\hat c-1)\bigr)}\times\nonumber\\
&& \times   \frac{\prod_{\substack{(i,j)\in \scriptscriptstyle{Y}\\ i-\text{odd}}}
  \bigl(4\Delta_2+(i-1)+(j-1)(\hat c-1)\bigr)\prod_{\substack{(i,j)\in \scriptscriptstyle{Y}\\ i-\text{odd}}}\bigl(4\Delta_3+(i-1)+(j-1)(\hat c-1)\bigr)}
  {\prod_{\substack{(i,j)\in \scriptscriptstyle{Y}\\ i-\text{odd}}}  
     \bigl(4\Delta + (i-1) + (Y_i'-j)(\hat c-1)\bigr)
    \prod_{\!\!\!\!\!\!\substack{(i,j)\in \scriptscriptstyle{Y}\\ (Y_j-i)-\text{odd}}} \!\!\!  \!\!
    \bigl(\hat c-(i-Y_j)-(Y_i'-j+1)(\hat c-1)\bigr)}\,.\nonumber
 \end{eqnarray}
Here we write $Y_j$ and $Y_i'$ for the numbers of boxes in the $j$th row and the $i$th column correspondingly. 
Below we list the results for the first two levels.

At level 1, we have $k_{1,2}=1$. There is only one such diagram $Y = \tableau{2}$. The corresponding contribution is
\begin{eqnarray}
\mathcal{B}_Y(\Delta_2,\Delta_3,\Delta)=\frac{2 \Delta_2 \Delta_3 (\hat c  - 4 \Delta - 1)}{(\hat c - 1 ) \Delta}\,.
\end{eqnarray}

At level 2, we have $k_{1,2}=2$. One finds two possible diagrams: $Y_1 = \tableau{4}$ and $Y_2 = \tableau{2 2}$. The corresponding contributions are
\begin{eqnarray}
&&\mathcal{B}_{Y_1}(\Delta_2,\Delta_3,\Delta)=
\frac{\Delta_2 \Delta_3(2 \Delta_2 +1) (2 \Delta_3 +1) (\hat c - 4 \Delta -3) (\hat c - 4 \Delta -1)}
{(\hat c -3)(\hat c -1) \Delta (2 \Delta +1)}\,,\nonumber\\
&&\mathcal{B}_{Y_2}(\Delta_2,\Delta_3,\Delta)=
\frac{4 \Delta_2\Delta_3 (\hat c + 4 \Delta_2 - 1)  (\hat c + 4 \Delta_3 - 1) (\hat c - 4 \Delta - 1)}
{(\hat c - 3) (\hat c - 1)^2 (\hat c + 4 \Delta - 1)}\,.
\end{eqnarray}

\newpage
%%%%%%%%%%%%%%%%%%%%%%%%%%%%%%%%%%%%%%%%%%%%%%%%%%%%%%%%%%%%%%%%%%%%%%%%%%%%%%%
\section*{Acknowledgements}
%%%%%%%%%%%%%%%%%%%%%%%%%%%%%%%%%%%%%%%%%%%%%%%%%%%%%%%%%%%%%%%%%%%%%%%%%%%%%%%
The author is grateful to
Marian Stanishkov, Yuji Tachikawa and Niclas Wyllard for useful discussions and comments.
The work was supported by the Russian Ministery of Education and
Science under the grants 2012-1.1-12-000-1011-012 and  
2012-1.1-12-000-1011-016.
The research has received funding from RFBR grant No.12-02-01092 and 
from the European Community`s Seventh Framework Programme EP7/2007-2013.

%%%%%%%%%%%%%%%%%%%%%%%%%%%%%%%%%%%%%%%%%%%%%%%%%%%%%%%%%%%%%%%%%%%%%%%%%%%%%%%
\providecommand{\href}[2]{#2}\begingroup\raggedright\endgroup

\end{document}